\documentclass{IEEEcsmag}
\IEEEoverridecommandlockouts
\usepackage{flushend}
\usepackage{caption}
\usepackage{subfigure}
\usepackage{subcaption}
\usepackage{graphicx}
\usepackage[table, svgnames, dvipsnames]{xcolor}
\usepackage{booktabs}
\usepackage{comment}
\usepackage{cite}
\usepackage{amsmath,amssymb,amsfonts}
\usepackage{algorithmic}
\usepackage{textcomp,balance,url}
\usepackage{multirow}
\usepackage{graphicx,array}

\usepackage{ragged2e}
\newcolumntype{C}[1]{>{\centering\let\newline\\\arraybackslash\parindent=0pt\hspace{0pt}}m{#1}}
\newcolumntype{L}[1]{>{\justifying\let\newline\\\arraybackslash\parindent=0pt\hspace{0pt}}m{#1}}
\newcolumntype{T}[1]{>{\raggedright\let\newline\\\arraybackslash\parindent=0pt\hspace{0pt}}m{#1}}
\usepackage[most]{tcolorbox} 
\definecolor{chestnut}{rgb}{0.97, 0.51, 0.47}
\def\BibTeX{{\rm B\kern-.05em{\sc i\kern-.025em b}\kern-.08em
    T\kern-.1667em\lower.7ex\hbox{E}\kern-.125emX}}
\usepackage{amsmath,amsfonts}
\usepackage{subfigure}
\usepackage{adjustbox}
\usepackage{comment}
\usepackage{soul}
\usepackage{textcomp}
\usepackage{siunitx}
\usepackage{booktabs}
\usepackage{stfloats}
\usepackage{xcolor}
\usepackage[most]{tcolorbox}
\usepackage{float}
\usepackage[verbatim]{minted}
\definecolor{lightboxblue}{RGB}{200,234,243}
\definecolor{lightboxgreen}{RGB}{222,241,204}
\definecolor{lightboxpink}{RGB}{252,235,238}
\usepackage{url}
\usepackage{verbatim}
\let\labelindent\relax
\usepackage{enumitem}
\usepackage{multirow}
\definecolor{IEEEblue}{RGB}{0,102,153}

\usepackage{ragged2e}
\usepackage{cite}
\usepackage{float}

\usepackage[table]{xcolor}
\usepackage{tikz}
\usepackage[dvipsnames]{xcolor}
\usetikzlibrary{arrows.meta, fit, backgrounds, positioning}
\definecolor{stageblue}{RGB}{0,45,120}
\definecolor{stagegreen}{RGB}{0,120,35}
\definecolor{stageorange}{RGB}{230,95,0}

\graphicspath{{./images/}}
\jmonth{November}
\jname{IEEE Internet Computing}
\jtitle{\emph{AeroPoint}: Agentic AI for Intent-Driven Point Cloud Inspection in UAV-Edge Systems}

\pubyear{2026}

\begin{document}
\sptitle{IEEE Internet Computing} 
\title{\emph{AeroPoint}: Agentic AI for Intent-Driven Point Cloud Inspection in UAV-Edge Systems}
\author{Reza Farahani*}
\affil{Distributed Systems Group, TU Wien, Austria}
\author{Zoha Azimi}
\affil{University of Klagenfurt, Austria}
\author{Naser Hossein Motlagh}
\affil{University of Helsinki, Finland}
\author{Christian Timmerer}
\affil{University of Klagenfurt, Austria}
\author{Arda Goknil}
\affil{Sintef, Norway}
\author{Sagar Sen}
\affil{Sintef, Norway}
\author{Schahram Dustdar}
\affil{Distributed Systems Group, TU Wien, Austria\\[4pt]
\small $^{\ast}$ Corresponding author: r.farahani@dsg.tuwien.ac.at
}
\markboth{IEEE Internet Computing}{IEEE Internet Computing}
\begin{abstract} \justifying
Unmanned Aerial Vehicle (UAV)-based Light Detection and Ranging (LiDAR) inspection generates dense point clouds that strain onboard computing and wireless bandwidth. Intent-based inspection allows operators to request objects, attributes, and spatial relationships through natural language. Answering these requests requires translating intent into spatial queries and retrieving relevant evidence under UAV processing and communication constraints. This article presents \textit{AeroPoint}, an agentic AI system combining onboard point cloud processing and knowledge graph construction with Large Language Model (LLM)-based retrieval planning at a ground edge server. The LLM translates inspection questions into structured requests, which the UAV executes to return selected graph evidence for answer generation. We evaluate the prototype in a power line inspection case study using \num{150} TS40K UAV LiDAR scenes and profile processing costs on NVIDIA Jetson Nano hardware. Retrieval experiments use graphs constructed from dataset-provided ground-truth semantic labels. Experimental results show that transmitted payload decreases from \qty{3700}{KB} for complete point clouds to \qty{3.1}{KB} for retrieval responses, while evaluated LLM routers achieve the highest graph retrieval completeness for relational queries.
\end{abstract}
\maketitle
\section{Introduction}
\label{sec:Introduction }
UAV-assisted inspection using onboard optical cameras provides detailed visual observations of hazardous and hard-to-access assets, including bridges, industrial facilities, and transport infrastructure. However, two-dimensional images lack explicit depth information, limiting direct geometric measurements~\cite{guan2021uav}. LiDAR complements visual sensing with three-dimensional point clouds whose spatial coordinates enable measurements of object dimensions, distances, and clearances, supporting inspection tasks that depend on spatial relationships among scene objects. Extracting this information requires operations such as segmentation, object identification, and spatial analysis on dense point clouds, placing substantial demands on limited onboard computing, memory, and battery resources.
Transmitting complete cloud data to ground servers, on the other hand, shifts the processing workload away from the UAV but incurs communication overhead and delays in inspection feedback. Interactive inspection thus requires selecting the scene information needed for each assessment before transmitting it across the UAV-edge link.

Recent work on intent-based UAV inspection allows operators to request information from captured images through natural language~\cite{farahani2026pixels,farahani2026evlm}. Extending this capability to LiDAR requires the UAV to organize dense point clouds into a compact representation of objects, attributes, and spatial relationships, making these scene elements directly queryable and allowing each inspection intent to determine which information crosses the UAV-edge link. Supporting this interaction also requires interpreting follow-up requests and reusing previously retrieved information. These requirements lead to our central research question: \textit{How can UAV and edge resources combine a compact onboard scene representation with intent-based retrieval to answer inspection requests while limiting processing and communication overhead?}

This article presents \textit{AeroPoint}, an agentic AI system for intent-driven point cloud inspection in UAV-edge systems. \textit{AeroPoint} integrates onboard segmentation and object identification to construct a knowledge graph of object instances, attributes, and spatial relationships. An LLM at the ground edge server translates inspection intents into structured retrieval requests. The UAV executes these requests and returns selected graph elements, which the LLM uses to generate an answer. Session storage preserves interaction history and retrieved evidence, supporting follow-up requests and avoiding redundant data transfers. 

We implement \textit{AeroPoint} on a UAV-edge testbed comprising an NVIDIA Jetson and an Intel Xeon Gold ground server, and evaluate it using \num{150} TS40K~\cite{lavado2024ts40k} UAV LiDAR scenes ia a power line inspection case study. The server hosts Qwen2.5-7B-Instruct as prompt-guided retrieval routers, evaluated on graphs constructed from dataset-provided semantic labels. Experimental results show that \textit{AeroPoint} reduces transmitted data by \qtyrange{97}{99}{\percent} compared with complete, sampled, and segmented point cloud transmission, and lowers total latency by \qtyrange{92}{98}{\percent} relative to complete-cloud transmission. Onboard power ranges from approximately \qtyrange{5}{12}{W}, with graph construction and retrieval adding little latency or energy beyond segmentation. LLM router achieves \qty{99}{\percent} graph-edge retrieval accuracy for relational queries.

\section{Related Work}
This section reviews related work in three categories and highlights \textit{AeroPoint}'s contributions.

\textbf{1) Offline point cloud analysis.}
Offline inspection pipelines combine UAV-acquired LiDAR and RGB imagery for 3D reconstruction, object detection, and defect localization. Guan et al.~\cite{guan2021uav} presented a LiDAR-assisted offline trajectory planning for power line inspection. Bonnin et al.~\cite{bonnin2026flight} combined onboard LiDAR navigation with post-flight reconstruction of vessel structures and image-based defect localization on the resulting 3D models. Li et al.~\cite{li2024high} fused images with LiDAR-derived depth information to detect bridge damage, distinguish defects from visual artifacts, and localize the findings in panoramic images. Luo et al.~\cite{luo2021detection} applied deep learning to UAV-acquired forest point clouds for individual tree detection. 
\begin{tcolorbox}[
    enhanced,
    width=\columnwidth,
    colback=gray!10,
    colframe=black,
    coltext=black,
    opacityback=1,
    opacityframe=1,
    fontupper=\small,
    boxrule=0.8pt,
    arc=3pt,
    outer arc=3pt,
    left=6pt,
    right=6pt,
    top=4pt,
    bottom=4pt]
\textbf{Limitations:} 
Despite advances in autonomous flight planning and multimodal perception, they rely on post-flight analysis to generate inspection findings. This delays semantic feedback and limits the ability to assess detected anomalies and adjust inspection activities during flight.
\end{tcolorbox}

\textbf{2) Real-time onboard perception.}
Onboard perception processes sensor observations during acquisition to support timely detection and assessment. Bultmann et al.~\cite{bultmann2023real} fused LiDAR, RGB, and thermal data through onboard segmentation and detection models. The resulting 3D semantic maps support object tracking, localization, and navigation, while label propagation provides training data for cross-domain adaptation. Pu et al.~\cite{pu2019real} used geometric analysis for onboard power line clearance assessment and communicated detected risks through lightweight 4G reports. Their framework also uses a 3D convolutional neural network to recognize tower configurations. Wang et al.~\cite{wang2023real} introduced 3D-ARSS, using sparse tensors and attention modules for LiDAR semantic segmentation on the onboard NVIDIA Jetson.
\begin{tcolorbox}[
    enhanced,
    width=\columnwidth,
    colback=gray!10,
    colframe=black,
    coltext=black,
    opacityback=1,
    opacityframe=1,
    fontupper=\small,
    boxrule=0.8pt,
    arc=3pt,
    outer arc=3pt,
    left=6pt,
    right=6pt,
    top=4pt,
    bottom=4pt]
\textbf{Limitations:} 
Although these approaches enable fast perception and hazard detection, they primarily support predefined perception and navigation tasks. Translating their outputs into context-aware inspection actions, such as revisiting suspicious areas or adjusting flight trajectories, requires further integration of semantic reasoning and decision-making.
\end{tcolorbox}

\textbf{3) Scene graphs and LLM-based spatial Reasoning.}
Scene graphs represent objects and their spatial and functional relationships~\cite{feng20233d,zhang2021knowledge,qiu20233d}, providing structured context for LLM-based reasoning. Yin et al.~\cite{yin2024sg} used hierarchical 3D scene graphs to support LLM-based, zero-shot localization of open-vocabulary navigation targets. Koch et al.~\cite{koch2024open3dsg} transferred vision-language features into a 3D graph neural network, enabling queryable object representations and open-vocabulary relationship descriptions from point clouds. Ranasinghe et al.~\cite{ranasinghe20263d} combined point cloud semantic perception, temporal scene graphs, and graph retrieval-augmented generation (GraphRAG) for contextual hazard reasoning via current and historical observations.
\begin{tcolorbox}[
    enhanced,
    width=\columnwidth,
    colback=gray!10,
    colframe=black,
    coltext=black,
    opacityback=1,
    opacityframe=1,
    fontupper=\small,
    boxrule=0.8pt,
    arc=3pt,
    outer arc=3pt,
    left=6pt,
    right=6pt,
    top=4pt,
    bottom=4pt]
\textbf{Limitations:} 
Despite enabling context-aware reasoning, graph processing and LLM inference impose computational and memory demands that complicate onboard UAV deployment. Existing work addresses ground-based navigation, scene querying, and hazard monitoring. Their integration into intent-driven UAV-edge inspection remains insufficiently explored, particularly when retrieving sufficient spatial evidence must satisfy onboard resource limits and wireless bandwidth constraints.
\end{tcolorbox}
\textbf{4) \textit{AeroPoint}'s Contribution.}
\textit{AeroPoint} makes dense UAV point clouds selectively queryable through compact onboard knowledge graphs of objects, attributes, and spatial relationships. A ground edge-hosted LLM agent requests intent-relevant graph evidence, allowing the UAV to transmit selected subgraphs instead of complete point clouds. Session reuse preserves follow-up context and reduces redundant transfers.

\section{\textit{AeroPoint} Design}
\label{sec:Framework}
\begin{figure*}
    \centering
    \includegraphics[width=1\linewidth]{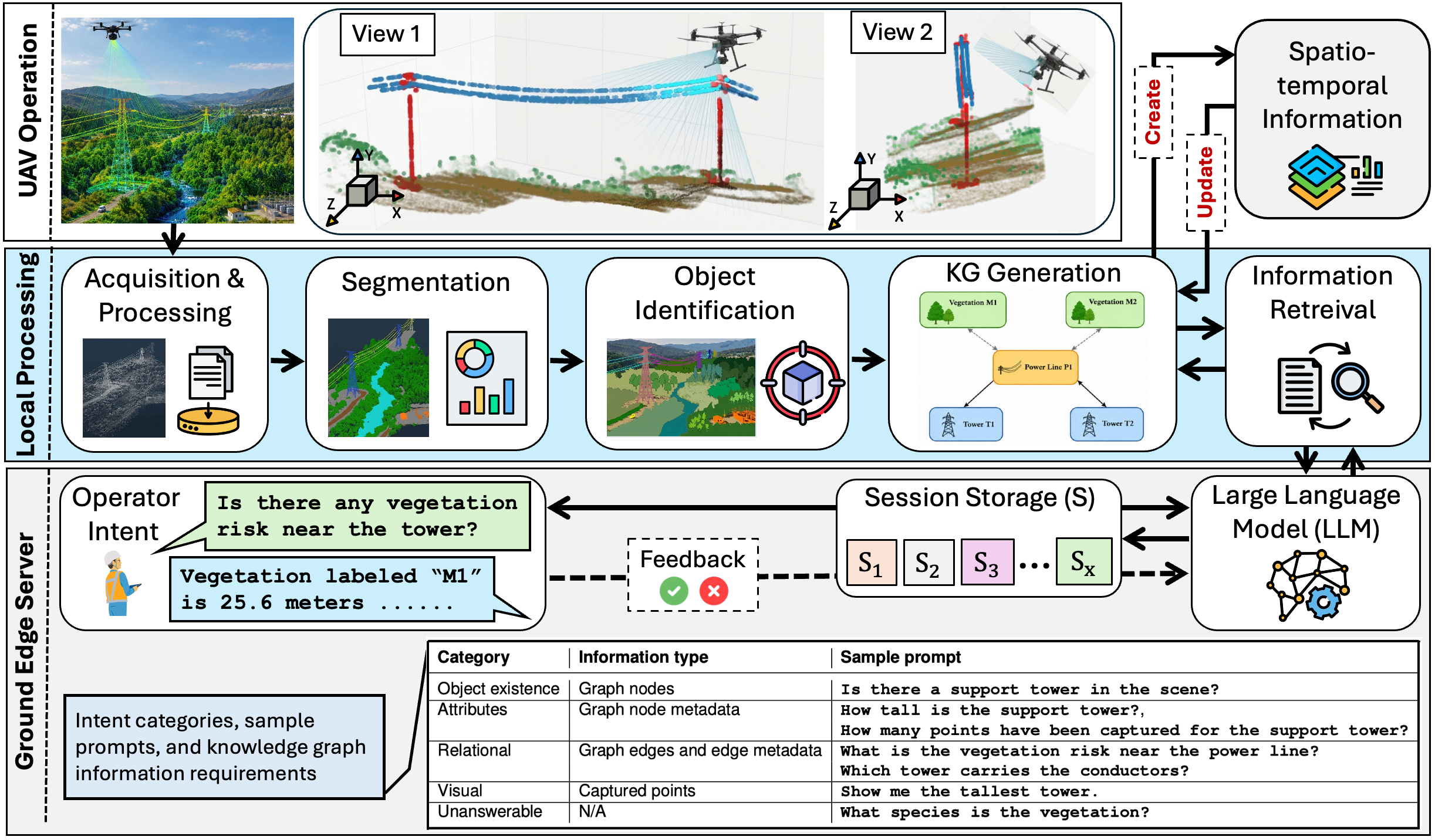}
     \caption{\emph{AeroPoint} system architecture. Spatio-temporal create/update operations maintain scene information, while session storage and operator feedback support inspection interactions. The table maps intent categories to their information requirements.}
    \label{fig:arch}
\end{figure*}

Fig.~\ref{fig:arch} organizes the \emph{AeroPoint} system architecture into three layers: \emph{UAV operation}, \emph{local processing}, and \emph{ground edge server}. These layers connect LiDAR acquisition to onboard scene representation and intent-based retrieval. Each captured scene is processed into a knowledge graph, enabling the ground edge server to request relevant objects and relationships rather than receive the complete point cloud.
\subsection{UAV Operation Layer}
The UAV captures LiDAR point clouds from different inspection viewpoints, with each point described by spatial coordinates $(x,y,z)$. For example, \emph{View~1} clearly distinguishes two towers and their connecting conductors. In \emph{View~2}, the changed viewpoint provides a partial observation in which the two towers are not clearly separated. The same infrastructure can therefore yield different point distributions and different evidence for object identification and relationship extraction. As further scans are acquired over time, the local processing layer must incorporate this additional or revised evidence into its scene representation.
\subsection{Local Processing Layer}
The local processing layer, illustrated in the middle of Fig.~\ref{fig:arch}, converts captured points into identifiable objects and relationships. Five components construct the onboard knowledge graph and support selective information retrieval.

\textbf{1) Acquisition and processing:}
This component receives the LiDAR point clouds and performs the initial processing required to prepare each captured scene for segmentation.

\textbf{2) Segmentation:}
Semantic segmentation assigns each point a label from a predefined class set. In the power line example, these labels distinguish towers, conductors, vegetation, and other scene elements. The labeled points are then passed to instance extraction to separate individual objects within each class.

\textbf{3) Object identification:}
Following segmentation, a clustering algorithm groups points within each semantic class into instances corresponding to individual physical objects. Each instance receives a unique identifier, while its semantic label, point count, center coordinates, and extreme-point coordinates are recorded as metadata. In the power line example, individual towers, power line segments, and vegetation clusters become distinct entities for graph construction.

\textbf{4) Knowledge graph generation:} 
The identified instances become graph nodes, with their metadata stored as node attributes. An ontology defines entity classes, attributes, and admissible relationship types in the architectural model. Candidate relationships between compatible instances are evaluated using spatial measurements and domain-specific criteria. A relationship becomes an edge when these criteria are satisfied, with the corresponding measurements stored as edge attributes.

For example, tower \texttt{T1}, conductor \texttt{P1}, and vegetation \texttt{M1} in Fig.~\ref{fig:arch} are represented as separate nodes. A \emph{support} edge connects \texttt{T1} to \texttt{P1} when their spatial configuration satisfies the support criteria. A \emph{clearance} edge connects \texttt{P1} to \texttt{M1} when their separation satisfies the proximity criteria, retaining the computed distance as an edge attribute. The graph thus exposes both the individual objects and the relationships needed to interpret their spatial context. These construction stages are repeated for each captured scene. The viewpoint-dependent observations described above can reveal additional instances or provide revised information about existing objects and relationships. The \emph{Create} and \emph{Update} operations in Fig.~\ref{fig:arch} incorporate this evidence into the spatio-temporal scene representation. For example, identifying another vegetation instance, \texttt{M2}, creates a new node and any qualifying relationships. Further observations of an existing instance can update its geometric attributes or the measurements associated with its edges.

\textbf{5) Information retrieval:} 
The final component of the local processing layer serves retrieval requests from the ground edge server. It extracts the relevant nodes, edges, and associated metadata from the onboard knowledge graph. For example, the ground edge LLM router translates ``Which vegetation is near conductor P1?'' into a request for \texttt{P1}, its associated vegetation nodes, and their \emph{clearance} edges with stored distances. These elements form a query-relevant subgraph, which the UAV returns to the LLM for answer generation instead of transmitting the complete point cloud or knowledge graph.

\subsection{Ground Edge Server Layer}
The ground edge server connects operator intents to the evidence maintained onboard. Its LLM performs two roles: routing requests to relevant scene information and generating answers from the retrieved evidence. Session storage preserves interaction context, while feedback records operator assessments.

\textbf{1) Operator prompts and intent categories:}
An inspection intent specifies what the operator wants to know about the observed scene, expressed through a natural-language request.

\textbf{2) LLM router:}
It translates the current prompt into a retrieval request specifying the relevant objects, relationships, and metadata. For example, ``Is there any vegetation risk near the tower?'' requires the relevant tower and vegetation nodes, together with their clearance relationships and stored distances. The router requests this evidence from the UAV's information retrieval component. Session context allows the request to refer to objects identified in preceding interactions.

\textbf{3) LLM answering:}
The LLM uses the returned subgraph and its metadata to formulate a natural-language answer. For the vegetation-proximity example, the response identifies the relevant vegetation instance, such as \texttt{M1}, and reports its stored clearance from the infrastructure. Object identifiers, attributes, and relationship measurements provide the evidence for the response. Requests for visual information additionally retrieve the captured points associated with the selected objects.

\textbf{4) Session storage:}
Session storage maintains the prompts and retrieved information associated with each UAV-operation session, represented by $\{S_1,S_2,\ldots,S_x\}$ in Fig.~\ref{fig:arch}. This context supports follow-up interpretation and reuse of previously retrieved graph elements. After an answer identifying \texttt{M1}, for example, ``How far is it from the power line?'' can refer to that vegetation instance. Its cached metadata can be reused while any additional clearance evidence is requested from the UAV, avoiding repeated transmission of the same object information.

\textbf{5) Feedback:}
Feedback keeps the operator more in the inspection loop by allowing them to assess whether a response addresses their intent and refine subsequent requests when clarification or additional evidence is needed. These assessments can be retained to evaluate response usefulness and guide improvements to the interaction.

\begin{table*}[!t]
\caption{Representative TS40K scene types and an example multi-turn inspection session. Operator questions are translated into JSON retrieval requests, selected subgraphs, and LLM responses, progressing from conductor existence to vegetation clearance near conductors and supporting towers.}
\label{tab:Table-1}
\centering
    \begin{tabular}{| L{6.2cm}  C{4.8cm} L{4.5cm}|}
    \hline
 \rowcolor{gray!24} \multicolumn{3}{|l|}{~~~~~~~~~~~~~~~~~~~~\textbf{No Tower}~~~~~~~~~~~~~~~~~~~~~~~~~~~~~~~~~~~~ \textbf{Tower Radius}~~~~~~~~~~~~~~~~~~~~~~~~~~~~~~~~~~~~~ \textbf{Power Line} } \\
\hline
\hline

 \multicolumn{3}{|c|}{
 \includegraphics[width=\linewidth]{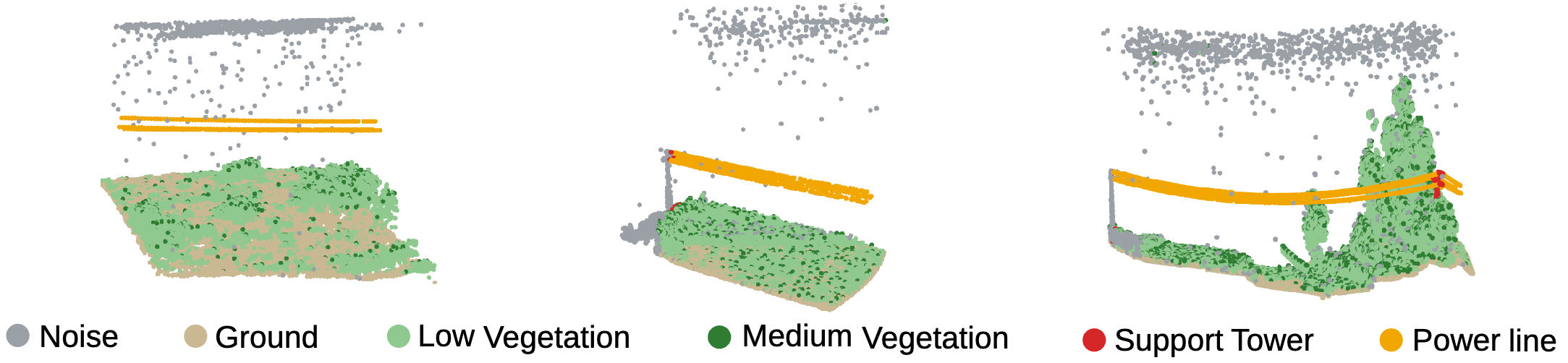}} \\
\midrule
\hline
  \rowcolor{gray!24}
\textbf{User Prompt \& Router Output} & \textbf{Retrieval Output} & \textbf{LLM Response} 
\\
\hline
\hline

\texttt{\textbf{Are there any power lines here?}} 

\begin{minted}[fontsize=\footnotesize,
               breaklines,               
               bgcolor=white!95!black,
               baselinestretch=1.1,
               xleftmargin=2mm]{json}
{"answerable": true,
 "node_ids":[],
 "node_classes":["conductor"],
 "edge_types": [],
 "hops": 0,
 "reason": "check for conductor nodes"}
\end{minted} 
& 
\raisebox{-0.5\height}{%
\begin{tikzpicture}[
    component/.style={
        draw,
        rounded corners=2pt,
        minimum width=1.2cm,
        minimum height=0.5cm,
        align=center,
        font=\scriptsize,
        inner sep=3pt
    }
]
\node[component, fill=Dandelion] (P1) {Power Line P1};

\end{tikzpicture}%
} 
&
 \texttt{Yes, power line is observed in the scene.}
\\  

\texttt{\textbf{What is the vegetation risk near P1?}} 
\begin{minted}[fontsize=\footnotesize,
               breaklines,
               bgcolor=white!95!black,
               baselinestretch=1.1,
               xleftmargin=2mm]{json}
{"answerable": true,
 "node_ids":["P1"],
 "node_classes":["vegetation", "conductor"],
 "edge_types": ["clearance"],
 "hops": 1,
 "reason": "clearance edges indicate risk"}
\end{minted}
& 
\begin{tikzpicture}[
    component/.style={
        draw,
        rounded corners=2pt,
        minimum width=1.1cm,
        minimum height=0.45cm,
        align=center,
        font=\tiny,
        inner sep=2pt
    },
    proximity/.style={
        <->,
        dashed,
        thin
    },
    distlabel/.style={
        font=\tiny,
        fill=white,
        inner sep=1pt
    }
]

\node[component, fill=YellowGreen] (M1) at (0,1.0) {Vegetation M1};
\node[component, fill=Dandelion] (P1) at (1.5,0) {Power Line P1};
\node[component, fill=YellowGreen] (M2) at (3.0,1.0) {Vegetation M2};

\draw[proximity]
    (M1) -- node[distlabel, sloped, above] {$d_{M1,P1}$} (P1);

\draw[proximity]
    (M2) -- node[distlabel, sloped, above] {$d_{M2,P1}$} (P1);

\end{tikzpicture}
&
 \texttt{Vegetation labeled as `M1` has a clearance distance of 25.6 meters from the conductor `P1`, which is within the typical safety margin for power lines. Vegetation labeled as `M2` has a clearance distance of 18.4 meters, also indicating potential
  risk. 
  } 
  \newline\\ 
\texttt{\textbf{What is the vegetation risk near the tower supporting P1?}}
\begin{minted}[fontsize=\footnotesize,
               breaklines,
               bgcolor=white!95!black,
               baselinestretch=1.1,
               xleftmargin=2mm]{json}
{"answerable": true,
 "node_ids":["P1"],
 "node_classes":["vegetation", "conductor", "tower"],
 "edge_types": ["clearance", "support"],
 "hops": 1,
 "reason": "check for both support and clearance edges."}
\end{minted}
&
\begin{tikzpicture}[
    component/.style={
        draw,
        rounded corners=2pt,
        minimum width=1.1cm,
        minimum height=0.45cm,
        align=center,
        font=\tiny,
        inner sep=2pt
    },
    proximity/.style={
        <->,
        dashed,
        thin
    },
    distlabel/.style={
        font=\tiny,
        fill=white,
        inner sep=1pt
    }
]

\node[component, fill=YellowGreen] (M1) at (0,1.0) {Vegetation M1};
\node[component, fill=Dandelion] (P1) at (1.5,0) {Power Line P1};
\node[component, fill=YellowGreen] (M2) at (3.0,1.0) {Vegetation M2};

\node[component, fill=RedOrange] (T1) at (0,-1.2) {Tower T1};
\node[component, fill=RedOrange] (T2) at (3.0,-1.2) {Tower T2};

\node[component, fill=YellowGreen] (M3) at (0,-2.3) {Vegetation M3};

\draw[proximity]
    (M1) -- node[distlabel, sloped, above] {$d_{M1,P1}$} (P1);

\draw[proximity]
    (M2) -- node[distlabel, sloped, above] {$d_{M2,P1}$} (P1);

\draw[proximity]
    (T1) -- node[distlabel, sloped, above] {$d_{T1,P1}$} (P1);

\draw[proximity]
    (T2) -- node[distlabel, sloped, above] {$d_{T2,P1}$} (P1);

\draw[proximity]
    (M3) -- node[distlabel, above] {$d_{M3,T1}$} (T1);

\end{tikzpicture}
& 
\texttt{Vegetation labeled as `M3` has a clearance distance of 28 meters from the tower `T1`, which is within the typical safety margin for transmission towers.  
  } 
\newline
\\
\\
\bottomrule
\end{tabular}
\end{table*}

\section{Evaluation Setup}
This section describes the \emph{AeroPoint} testbed and experimental configuration.

\textbf{1) UAV-edge hardware configuration:}
We run the UAV-side pipeline on an NVIDIA Jetson Nano device with an \num{8}-core CPU, \qty{16}{GB} RAM, \num{1024} CUDA cores, and \num{32} Tensor cores. Ground edge processing runs on an Intel Xeon Gold server with \num{128} CPU cores and two NVIDIA Quadro GV100 GPUs.

\textbf{2) Point cloud dataset:} We used UAV-mounted LiDAR scans from TS40K~\cite{lavado2024ts40k}, capturing rural terrain and the electrical transmission network from a bird's-eye perspective. The dataset spans three scene types: \emph{(i) Tower-radius}, focusing on supporting towers and their environment; \emph{(ii) Power line}, focusing on power lines with two supporting towers; and \emph{(iii) No tower}, representing rural terrain containing the transmission network, excluding supporting towers but potentially including power lines. Table~\ref{tab:Table-1} shows one sample from each scene type. Each point is assigned one of six labels: Noise, Ground, Low Vegetation, Medium Vegetation, Power Line, or Support Tower.

We randomly selected \num{150} distinct scenes for evaluation, averaging \num{655600} points per scene. We then applied random sampling to the point clouds and compared segmentation execution time and the number of identified objects across sampling budgets. Samples containing 24K-49K points required \qtyrange{0.93}{1.13}{s} for segmentation and yielded an average of \numrange{2}{2.4} identified objects per scene. At 65K points, segmentation required \qty{1.2}{s} and yielded \num{3} objects, compared with \qty{1.6}{s} and \num{5.3} objects at 98K points. Full point clouds yielded an average of \num{5.8} objects. We selected 65K points as a trade-off between segmentation execution time and object identification.

\textbf{3) Segmentation:}
For benchmarking, we selected PointNet++~\cite{qi2017pointnet++}, one of the architectures evaluated in the original TS40K study~\cite{lavado2024ts40k}. Our evaluation workflow used the dataset-provided ground truth labels without retraining or fine-tuning PointNet++ on TS40K. Separately, to assess deployment feasibility on edge devices, we implemented the same PointNet++ architecture with randomly initialized weights. This configuration was used exclusively to measure inference-time resource requirements, including computational and memory usage, and was not used to evaluate classification performance.

\textbf{4) Object identification:}
We used the ground truth semantic labels from TS40K to identify individual objects in each point cloud. Tower, conductor, and vegetation points were processed independently using Density-Based Spatial Clustering of Applications with Noise (DBSCAN)~\cite{ester1996density}, with class-specific distance thresholds and minimum cluster sizes. Clusters below the minimum point requirements were discarded. For each retained instance, we computed its centroid, spatial extent, height, and vertical extent. Object heights were estimated relative to the nearest ground points in the horizontal plane to account for terrain variation, particularly on slopes. TS40K point clouds use normalized coordinates in $[0,1]$. We applied predefined scaling factors to the coordinates and derived geometric properties to obtain estimated metric values. These factors were not derived from field measurements or surveyed object dimensions. They provide an approximate physical scale for interpreting and using spatial information in subsequent LLM-based experiments.

\textbf{5) Knowledge graph generation:}
We generated the knowledge graph from the extracted object instances to represent structural connections between towers and conductors and the spatial proximity of vegetation to infrastructure. Each instance became a node containing its semantic class and geometric properties, including its centroid, dimensions, and height. Although the proposed architecture supports relationships defined through a domain-specific ontology, no such ontology was available for this work. We therefore used predefined geometric rules aligned with the monitoring objectives, evaluating relationships through spatial distances between the corresponding instance point clouds. A tower and a conductor were considered connected when the minimum distance between their point clouds was below \qty{2}{m}. Vegetation was associated with the corresponding infrastructure node through a \emph{clearance} relationship when located within \qty{30}{m} of a conductor or between \qty{1}{m} and \qty{30}{m} of a tower.

\textbf{6) Operator intents:}
We define five intent categories according to their evidence requirements, as summarized in the table of Fig.~\ref{fig:arch}: \emph{(i) Object-existence} queries require graph nodes, \emph{(ii) attribute} queries require node metadata such as height or point count, and \emph{(iii) relational} queries require edges and edge metadata. \emph{(iv) Visual} queries require relevant captured points, as in ``Show me the tallest tower''. \emph{(v) Unanswerable} queries concern information absent from the representation, such as vegetation species.
     

\textbf{7) LLM router:}
We evaluated \texttt{gemma-4-E4B-it} and \texttt{Qwen2.5-7B-Instruct} as retrieval routers. A system prompt defines the routing task and requires JSON requests specifying the graph evidence needed to answer each question. The retrieval interface remains independent of the underlying LLM. Table~\ref{tab:Table-1} shows representative scenes and sample user prompts, together with the JSON router outputs, retrieved subgraphs, and LLM responses. If the retrieval component finds no matching nodes or edges, it returns empty lists for the corresponding graph elements.
For \texttt{Are there any power lines here?}, the router generates a JSON request for \texttt{conductor} nodes without requesting edges, since nodes alone establish object existence. Once the retrieved conductor is identified as \texttt{P1}, the follow-up \texttt{What is the vegetation risk near P1?} anchors the request at \texttt{P1}, adds the \texttt{vegetation} node class, and requests \texttt{clearance} edges with their distance metadata. The final question about the tower supporting \texttt{P1} additionally requests the \texttt{tower} class and \texttt{support} edges, extending retrieval to the supporting infrastructure and nearby vegetation.
\fbox{\parbox{1\linewidth}{
\small
\textbf{Router system prompt:}
\texttt{You are a retrieval router for a LiDAR power line corridor inspection system. A scene graph exists on a remote UAV. Your job is to determine the minimum subgraph required to answer the user's question and emit a retrieval request in JSON. The request is sent over a bandwidth-constrained link, so every unnecessary byte has a cost.}
}}

\textbf{8) Retrieval:}
The retrieval module executes each JSON request against the onboard graph and returns the selected subgraph with its metadata. In the illustrated interaction in Fig.~\ref{fig:arch}, the first request returns \texttt{P1} and its attributes. The second adds vegetation nodes \texttt{M1} and \texttt{M2} and their clearance edges to \texttt{P1}. The third expands the evidence to towers \texttt{T1} and \texttt{T2}, vegetation \texttt{M3}, and the corresponding support and clearance relationships. Previously retrieved metadata, including \texttt{P1}, are reused from the session cache to avoid redundant transfers. When no matching nodes or edges are found, the corresponding lists are empty.

\textbf{9) Response generation:}
The LLM generates a natural-language answer using the retrieved UAV evidence and previously cached information. In Table~\ref{tab:Table-1}, the first response confirms conductor presence. Subsequent responses report the stored vegetation clearances: \qty{25.6}{m} and \qty{18.4}{m} from \texttt{M1} and \texttt{M2} to \texttt{P1}, respectively, followed by \qty{28}{m} from \texttt{M3} to tower \texttt{T1}.

\section{Evaluation Results} \label{sec:Results}
This section evaluates \textit{AeroPoint} retrieval accuracy, onboard processing costs, and the communication-latency trade-off.

\textbf{1) Retrieval accuracy:}
Retrieval accuracy measures the fraction of query-required graph nodes and edges retrieved by the router. For each query, we generated reference targets by applying its predefined intent to the corresponding scene graph. The intent specifies the node classes, edge types, and information type, following the categories in Fig.~\ref{fig:arch}. Category-specific rules determine the relevant instances, relationships, or graph traversal for that scene.

These targets serve exclusively as reference annotations for scoring and are not supplied to the inference pipeline. The expected evidence therefore adapts to each scene, while the evaluation rules remain consistent across the dataset. The resulting scores assess retrieval from existing graphs, assuming ground truth semantic labels and graph structures. Against these references, Table~\ref{tab:accuracy} shows complete node retrieval for object-existence queries with two different LLM routers. For relational queries, both achieve \qty{99}{\percent} edge completeness, while node completeness reaches \qty{72}{\percent} for Qwen and \qty{86}{\percent} for Gemma. This exposes a limitation in selecting all required object instances despite strong relationship coverage. Gemma also achieves higher attribute-node completeness, at \qty{98}{\percent} compared with Qwen's \qty{84}{\percent}. For visual intents, Qwen retrieves all required nodes, whereas Gemma reaches \qty{54}{\percent}, showing that retrieval performance depends on both the router and the requested evidence type.
\begin{table}[t!]
    \centering
    \caption{Node and edge retrieval accuracy across intent categories. Scores denote the fraction of required graph elements retrieved.}
    \label{tab:accuracy}
    \small
    \fontsize{7.5pt}{7.5pt}\selectfont    
    \begin{tabular}{lcccc}
        \toprule
        & \multicolumn{2}{c}{\textbf{Qwen}} 
        & \multicolumn{2}{c}{\textbf{Gemma 4}} \\
        \cmidrule(lr){2-3} \cmidrule(lr){4-5}
        \textbf{Category} 
        & \textbf{Nodes} & \textbf{Edges} 
        & \textbf{Nodes} & \textbf{Edges} \\
        \midrule
        Object existence & 1.00 & 1.00 & 1.00 & 1.00 \\
        Attributes        & 0.84 & 0.89 & 0.98 & 0.83 \\
        Relational        & 0.72 & 0.99 & 0.86 & 0.99 \\
        Visual            & 1.00 & 0.89 & 0.54 & 0.88 \\
        
        \bottomrule
    \end{tabular}
\end{table}

\begin{figure*}
    \centering
    \includegraphics[width=1\linewidth]{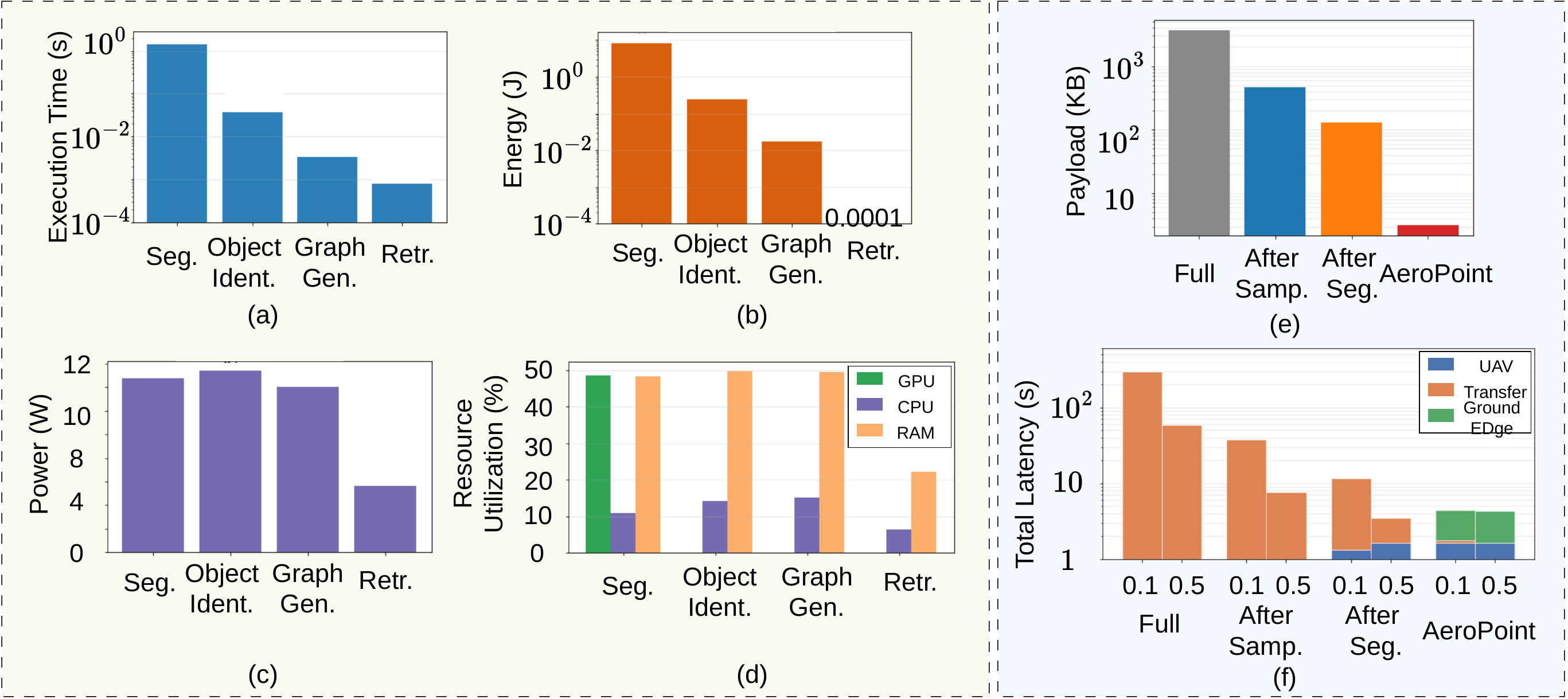}
    \caption{\emph{AeroPoint} processing and communication costs: (a) stage execution time, (b) energy per execution, (c) power draw, (d) GPU, CPU, and RAM utilization, (e) transmitted payload, and (f) reported latency comprising UAV processing, transfer, and ground edge router execution at uplink rates of \qty{0.1}{\mega\bit\per\second} and \qty{0.5}{\mega\bit\per\second}.}
    \label{fig:UAV}
\end{figure*}

\textbf{2) Onboard processing costs:}
We check the feasibility of onboard processing and profile its steps using execution time, energy per execution, power draw, and GPU, CPU, and RAM utilization. Fig.~\ref{fig:UAV}(a) shows that segmentation dominates execution time. Object identification operates on the order of tens of milliseconds, graph generation takes a few milliseconds, and retrieval remains below one millisecond. The energy profile in Fig.~\ref{fig:UAV}(b) follows the same ordering, i.e., segmentation consumes the most energy, while retrieval energy is reported as \qty{0.0001}{J}.

Fig.~\ref{fig:UAV}(c) shows power draw of approximately \qtyrange{11}{12}{W} during segmentation, object identification, and graph generation, compared with approximately \qty{5}{W} during retrieval. Although graph generation draws power comparable to segmentation, its short execution time keeps its energy contribution small. Resource utilization in Fig.~\ref{fig:UAV}(d) shows GPU activity concentrated in segmentation, at approximately \qty{48}{\percent}. Across the stages, CPU utilization is approximately \qtyrange{6}{15}{\percent}, while RAM utilization ranges from approximately \qtyrange{22}{50}{\percent}. These profiles show that graph construction and retrieval add limited processing and energy costs relative to the segmentation workload.

\textbf{3) Communication and latency}
We compare \emph{AeroPoint} with three transmission baselines: \emph{Full}, transmitting the complete point cloud; \emph{After sampling}, transmitting the sampled cloud; and \emph{After segmentation}, transmitting the segmented point data. \emph{AeroPoint} transmits the selected graph evidence.
Fig.~\ref{fig:UAV}(e) shows that \emph{AeroPoint} transmits \qty{3.1}{KB}, compared with \qty{3700}{KB}, \qty{472}{KB}, and \qty{129}{KB} for the three baselines, reductions of \qty{99}{\percent}, \qty{99}{\percent}, and \qty{97}{\percent}, respectively. In Fig.~\ref{fig:UAV}(f), the labels \texttt{0.1} and \texttt{0.5} denote transfer rates in megabits per second, with two bars per method. The stacked bars combine UAV processing, transfer, and router execution times. The full-cloud and sampled-cloud baseline totals include transfer time only. Transmission after segmentation includes \qty{1.39}{s} of UAV processing, while \emph{AeroPoint} includes \qty{1.43}{s} of UAV processing and \qty{2.77}{s} of router execution. 

At \qty{0.1}{\mega\bit\per\second}, transferring the \emph{AeroPoint} response takes \qty{0.248}{s}, giving a total of \qty{4.448}{s}. The baseline totals are \qty{296}{s}, \qty{37.76}{s}, and \qty{11.71}{s}. Selective retrieval therefore reduces total latency by \qty{98}{\percent}, \qty{88.2}{\percent}, and \qty{62.0}{\percent}, respectively. At \qty{0.5}{\mega\bit\per\second}, response transfer takes \qty{0.0496}{s}, reducing the \emph{AeroPoint} total to \qty{4.2496}{s}. Full-cloud and sampled-cloud transmission require \qty{59.2}{s} and \qty{7.552}{s}, respectively. However, transmission after segmentation requires \qty{3.454}{s}, approximately \qty{23}{\percent} less than the \emph{AeroPoint} total. The smaller payload provides its strongest latency benefit at the lower transfer rate, while router execution limits the benefit as transfer time decreases.
\begin{tcolorbox}[
    enhanced,
    width=\columnwidth,
    colback=gray!10,
    colframe=black,
    coltext=black,
    opacityback=1,
    opacityframe=1,
    fontupper=\small,
    boxrule=0.8pt,
    arc=3pt,
    outer arc=3pt,
    left=6pt,
    right=6pt,
    top=4pt,
    bottom=4pt]
\textbf{Result summary:}
\emph{AeroPoint} reduces payload by \qtyrange{97}{99}{\percent} with limited onboard graph-construction and retrieval overhead. LLM Routers achieve \qty{99}{\percent} relational-edge completeness, while object selection remains a retrieval limitation. Router execution limits latency gains at higher transfer rates, where segmented-cloud transmission can be faster.
\end{tcolorbox}

\section{Conclusion and Future Work}
This article presented \emph{AeroPoint}, an agentic AI system for intent-driven point cloud inspection across UAV and ground edge resources. It combines compact onboard knowledge graphs with LLM-based subgraph retrieval and session reuse, replacing complete point cloud transfers with query-relevant scene evidence. Evaluation on \num{150} TS40K scenes and on real UAV-edge devices demonstrated \qtyrange{97}{99}{\percent} payload reduction and \qty{99}{\percent} relational-edge completeness with both routers, while graph construction and retrieval added limited onboard overhead. These results support selective evidence exchange for interactive inspection, although latency gains depend on transfer bandwidth and router execution cost. Future work will evaluate the complete pipeline with predicted semantic labels and explore adaptive retrieval under varying bandwidth.
\section*{Acknowledgment}This work was supported by the FFG EdgeAI-Drone project (FO999944705).
\bibliographystyle{IEEEtran}
\bibliography{bibtex}
\newpage
\begin{IEEEbiographynophoto}{Reza Farahani}  
is a Univ.Ass. Postdoctoral Researcher at TU Wien and a Lecturer at the University of Klagenfurt, Austria, where he received his Ph.D. in Computer Science in 2023 and is completing his habilitation. He coordinates the FFG EdgeAI-Drone project and has contributed to EU ENFIELD, CDL ATHENA, and EU Graph-Massivizer, co-leading WP5 on serverless orchestration and large-scale edge-cloud testbeds. He has published 60+ scientific publications and received the IEEE HiPC 2024 Distinguished Paper Award. His research focuses on distributed and networked systems, serverless computing, distributed multimedia, agentic edge AI, and Green AI. Contact him at: r.farahani@dsg.tuwien.ac.at
\end{IEEEbiographynophoto}
\begin{IEEEbiographynophoto}{Zoha Azimi}  
is a Ph.D. candidate at the University of Klagenfurt, Austria. She received her M.Sc. in artificial intelligence from the University of Bologna, Italy. Her research interests include AI, energy-efficient multimedia systems, and agentic AI. Contact her at: zoha.azimi@aau.at
\end{IEEEbiographynophoto}
\begin{IEEEbiographynophoto}{Naser Hossein Motlagh} is a Senior Researcher at the University of Helsinki, Finland. He received his D.Sc. in Networking Technology at Aalto University, Finland, in 2018. His research interests include the Internet of Things, wireless sensor networks, edge AI, environmental sensing, and unmanned aerial vehicles. Contact him at: naser.motlagh@helsinki.fi
\end{IEEEbiographynophoto}
\begin{IEEEbiographynophoto}{Arda Goknil} 
received his Ph.D. degree in computer science from the University of Twente, in the Netherlands. He is a senior research scientist at SINTEF, Norway. He was a research associate at the University of Luxembourg. His research concerns AI Engineering, Sustainable AI, Software Testing, Software Security, and Intermittent Computing.
\end{IEEEbiographynophoto}
\begin{IEEEbiographynophoto}{Sagar Sen} is a Senior Research Scientist and Research Group Leader of Trustworthy Green IoT Software at SINTEF Digital, Norway. His research focuses on AI engineering, trustworthy and sustainable AI, software engineering, and cyber-physical systems. He received his PhD from Univ de Rennes 1/INRIA Rennes, France, in 2010, and previously worked as a postdoctoral researcher in France, a research scientist at Simula, and VP of Product at Sweetzpot. He has authored over 140 scientific publications.
\end{IEEEbiographynophoto}
\begin{IEEEbiographynophoto}{Christian Timmerer}
is a Full Professor with the University of Klagenfurt, Austria, and the Director of the Christian Doppler Laboratory ATHENA. He is also a Co-Founder and Chief Innovation Officer of Bitmovin. His research interests include multimedia systems, adaptive streaming, and drone communication. Contact him at: christian.timmerer@aau.at
\end{IEEEbiographynophoto}
\begin{IEEEbiographynophoto}{Schahram Dustdar}  
is a Full Professor of Computer Science and Head of the Distributed Systems Group at TU Wien, Austria. He is Chairperson of the Informatics Section of Academia Europaea and has served as an Associate Editor for several IEEE and ACM journals. His distinctions include the ACM Distinguished Scientist and IBM Faculty Awards. Contact him at: dustdar@dsg.tuwien.ac.at.
\end{IEEEbiographynophoto}
\end{document}